\documentclass[aps,pra,twocolumn,groupedaddress,10pt]{revtex4-1}
\usepackage{mathrsfs}
\usepackage{amsmath}
\usepackage{amsfonts}
\usepackage{amssymb}
\usepackage{amsthm}
\usepackage{graphicx}
\usepackage{color}
\usepackage{graphics}
\usepackage{epsfig}
\usepackage{ulem}
\usepackage{subfigure}
\providecommand{\U}[1]{\protect\rule{.1in}{.1in}}

\begin{document}
\title{Reconfigurable topological spin wave beamsplitters and interferometers}
\author{X. S. Wang$^{1,2}$}
\author{H. W. Zhang$^{1}$}
\author{X. R. Wang$^{2,3,}$}
\email[Corresponding author:]{phxwan@ust.hk}
\affiliation{$^1$School of Microelectronics and Solid-State Electronics,
University of Electronic Science and Technology of China, Chengdu,
Sichuan 610054, China}
\affiliation{$^2$Department of Physics, The Hong Kong University of
Science and Technology, Clear Water Bay, Kowloon, Hong Kong}
\affiliation{$^3$HKUST Shenzhen Research Institute, Shenzhen 518057, China}

\begin{abstract}
Conventional magnonic devices use three classes of magnetostatic waves
that require detailed manipulation of magnetization structure that makes
the design and the device/circuitry scalability difficult tasks. Here,
we demonstrate that devices based on topological exchange spin waves do
not suffer from the problem with additional nice features of nano-scale
wavelength and high frequency. Two results are reported. 1) A perpendicular
ferromagnet on a honeycomb lattice is generically a topological magnetic
material in the sense that topologically protected chiral edge spin waves
exist in the band gap as long as spin-orbit induced nearest-neighbor
pseudodipolar interaction (and/or next-nearest neighbor Dzyaloshinskii-Moriya
interaction) is present. 2) As a proof of concept, spin wave beamsplitters
and spin wave interferometers are designed by using domain walls to manipulate
the propagation of topologically protected chiral spin waves. Since magnetic
domain walls can be controlled by magnetic fields or electric current/fields,
one can essentially draw, erase and redraw different spin wave devices and
circuitry on the same magnetic plate so that the proposed devices are
reconfigurable and tunable. Devices made from magnetic topological materials
are robust against both internal and external perturbations such as the spin
wave frequency variation and device geometry as well as defects.
\end{abstract}

\maketitle

\section{INTRODUCTION}
Spintronics is about generation, detection and manipulation of spins for
information storage and processing. Similar to electron spintronics that
deals with electron spin, magnon spintronics, known also as magnonics
\cite{book1,magnonics1,magnonics2}, utilizes magnon spin that has the
advantages of low energy consumption and long coherence length
\cite{magnonics2,Kajiwara,Bauer}. Magnons, the quanta of spin waves,
are promising information carriers as well as control knob of spin
textures \cite{AMSTT,xiansi,entropy} that is the subject of intensive
researches in recent years. Various spin wave devices and circuits
such as logic gates \cite{gate1,gate2}, filters \cite{filter},
waveguides \cite{waveguide,bend}, diodes \cite{diode}, and
multiplexors \cite{multi}, have been proposed and designed.
The important functionality of these devices is to manipulate spin
waves and to control spin wave propagation in a designed way.
However, the performance of these devices are usually not stable
against internal and external perturbations such as spin wave frequency
variation or device geometry change. Interestingly, recently
discovered magnonic topological matters \cite{TM1,TM2,TM3,ours} have
topologically protected unidirectional spin waves that are well confined
on the sample surfaces and edges, and whose propagation is very robust
against internal and external perturbations, in contrast to the fragile
nature of conventional spin waves. Thus, devices based
on the topological spin waves should not suffer from the usual problems
of conventional spin wave devices.

\begin{figure}[htb]
\begin{center}
\includegraphics[width=8.5cm]{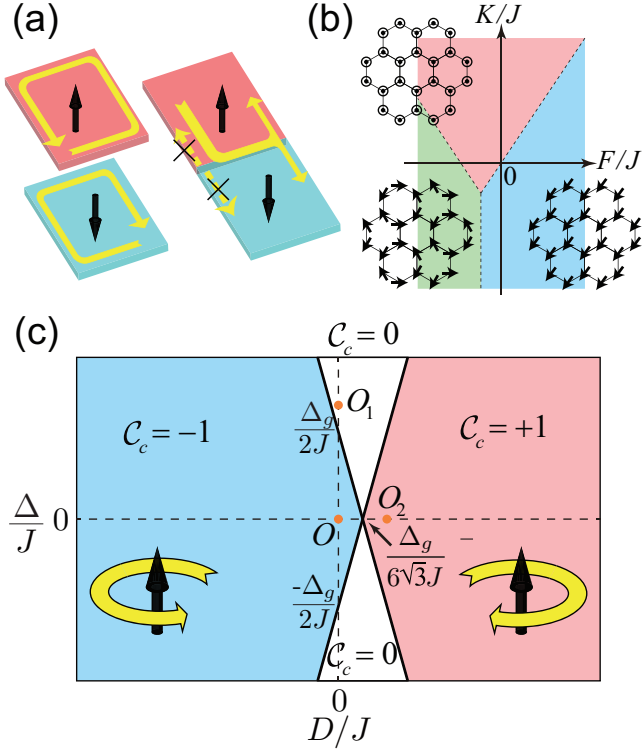}
\end{center}
\caption{
(a) Schematic illustration for the topological spin wave edge states
(left) and the domain wall SWBS (right). The red and cyan regions denote
domains in which spins point to the $+z$ and $-z$ directions, respectively.
The yellow arrows denote the spin wave propagation direction.
(b) Various phases in the $K/J-F/J$ plane when $D=\Delta=0$.
The spin arrangements in these phases are shown in the insets.
(c) Topological phase diagram in the $D/J-\Delta/J$ plane (for $K=J$ and $F=0.01J$).
The cyan (pink) region is topologically nontrivial phase with Chern number
$\mathcal{C}_c=-1$ ($\mathcal{C}_c=+1$) for the conduction band, and the
edge spin waves propagate counterclockwise (clockwise) with respect to
the magnetization direction, as illustrated in the insets.
The white regions are topologically trivial phase with $\mathcal{C}_c=0$. }
\label{model}
\end{figure}

A spin wave beamsplitter (SWBS) can split an incoming spin wave beam
into two or more outgoing beams. One existing SWBS is based on the
inter-conversion between magnetostatic surface spin waves and backward
volume magnetostatic spin waves in a T-junction \cite{splitter}.
However, the wavelength of magnetostatic spin waves is usually hundreds of
nanometers to millimeters long, an intrinsic problem for miniaturization.
Also, because these spin waves require specific configuration of the
sample and the external field \cite{splitter,MSW}, such a SWBS is
difficult to use, especially in constructing 1-to-$n$ splitter for $n>2$.
In this paper, we propose to use domain walls in topological magnetic film
as SWBSs and spin wave interferometers, two basic elements in magnonics.
The topological chiral edge spin waves propagate in a certain
direction with respect to the magnetization direction, as schematically
illustrated in the left panel of Fig. \ref{model}(a).
Consider an edge spin wave entering a domain wall that separates two
domains as shown in the right panel Fig. \ref{model}(a).
Since the topologically protected edge spin waves propagate in opposite
directions in the two domains, a spin wave propagating
towards the domain wall can neither penetrate it nor be reflected by it.
It must move along the domain wall. When the spin wave beam reaches the
other edge, it will split into two beams propagating in opposite directions.
Thus a domain wall is essentially a 1-to-2 SWBS for the topologically
protected spin waves in the bulk band gap.
We further show that inside the bulk band gap, there are two unidirectional
spin wave modes inside a domain wall for a given frequency due to two
topologically protected edge spin waves, one from each domain.
When a chiral edge spin wave of a given frequency from one domain propagates
towards a domain wall, two eigenmodes of the same frequency but different
wavenumbers are
excited inside the domain wall and propagate along the domain wall.
The overlap of the two spin waves results in an interference pattern inside
the domain wall. After the spin waves passing through the domain wall,
one beam splits into two, and the power division ratio depends on the
wavenumbers, group velocities of the two modes, and the domain wall length,
but not on the position of the domain wall or the wave source.
The idea can be generalized to 1-to-$n$ spin wave splitting.
A Mach-Zehnder-type spin wave interferometer is also designed.
A spin wave beam is first split into two and recombine later to form an
interference pattern that varies periodically with the relative phase
change of the two beams.

\section{MODEL AND ITS PHASE DIAGRAM}
We consider classical spins on a honeycomb lattice of lattice constant
$a$ in the $xy$ plane. The Hamiltonian is
\begin{multline}
\mathcal{H}=-\frac{J}{2}\sum_{\left\langle i,j \right\rangle} \mathbf{m}_i\cdot
\mathbf{m}_j-\frac{F}{2}\sum_{\left\langle i,j \right\rangle}
(\mathbf{m}_i\cdot \mathbf{e}_{ij})(\mathbf{m}_j\cdot \mathbf{e}_{ij})\\
-D\sum_{\langle\langle i,j\rangle\rangle}\nu_{ij}
\hat{\mathbf{z}}\cdot\left(\mathbf{m}_i\times\mathbf{m}_j\right)-\sum_{i}\frac{K_i}{2}m_{iz}^2,
\label{Hami}
\end{multline}
where the first term is the nearest-neighbor ferromagnetic Heisenberg
exchange interaction with $J>0$. The second and third terms arise from
the spin-orbit coupling (SOC) \cite{pseudo,DMI}. $\mathbf{e}_{ij}$ is
the unit vector pointing from site $i$ to $j$. The second term
is the nearest-neighbor pseudodipolar interaction of strength $F$,
which is the second-order effect of the SOC [the first-order effect, the
nearest-neighbor Dzyaloshinskii-Moriya interaction (DMI), vanishes because
the center of the A-B bond is an inversion center of the honeycomb lattice].
The third term is the next-nearest-neighbor DMI of strength $D$ with
$\nu_{ij}=\frac{2}{\sqrt{3}}\hat{\mathbf{z}}\cdot(\mathbf{e}_{li}
\times\mathbf{e}_{lj})=\pm 1$, where $l$ is the nearest neighbor site of $i$ and $j$.
The last term is the sublattice-dependent anisotropy whose easy-axis is
along the $z$ direction with anisotropy coefficients of $K_i=K+\Delta$
for $i\in$ A and $K-\Delta$ for $i\in$ B. $\mathbf{m}_i$ is the unit vector
of the magnetic moment of magnitude $\mu$ at site $i$. The dynamics of the
spins is governed by the Landau-Lifshitz-Gilbert (LLG) equation \cite{LLG,ours},
\begin{equation}
\frac{\mathrm{d}\mathbf{m}_i}{\mathrm{d} t}=-\gamma\mathbf{m}\times \mathbf{H}^
\text{eff}_i+\alpha \mathbf{m}_i\times \frac{\mathrm{d}\mathbf{m}_i}{\mathrm{d} t},
\label{LLG}
\end{equation}
where $\gamma$ is the gyromagnetic ratio and $\alpha$ is the Gilbert damping
constant. $\mathbf{H}^\text{eff}_i=\frac{\partial \mathcal{H}}{\mu_0\mu\partial
\mathbf{m}_i}$ is the effective field at site $i$.
Out of five model parameters in Hamiltonian \eqref{Hami}, $J$ can be the
natural energy unit. The natural units of the time, length and magnetic field
are $\frac{\mu_0\mu}{\gamma J}$, $a$ and $\frac{J}{\mu_0\mu}$.
The LLG equation is numerically solved by using a homemade \texttt{C++}
code with fourth-order Runge-Kutta method. We first determine the ground
state of the system by numerically relaxing the spins to their stable
state starting from an initial configuration in which the spins are randomly
and uniformly distributed in a cone with polar angle $\theta<15^\circ$.
For $D=0$ and $\Delta=0$, the phase diagram of the model in the $K/J-F/J$
plane is shown in Fig. \ref{model}(b). The system is in a ferromagnetic
topological phase \cite{ours} with spins aligning along the $z$ direction
(pink) when $3J+K>-\frac{3}{2}F$ and $K>\frac{3}{2}F$.
The system has a bulk band gap of $\Delta_g=(3J+K)-\sqrt{(3J+K)^2-9F^2/4}$
in the case of $F\neq 0$ in this phase with
Chern number $\mathcal{C}_c=-1$ for the conduction band and $\mathcal{C}_v=+1$
for the valence band, and topologically protected chiral edge spin waves in
the gap for a finite system. When $3J+K<-\frac{3}{2}F$ and $F<-J$(green),
the ground state has a chiral spin structure in which spins lie in the
$xy$ plane with zero net magnetic moment on each hexagon. As shown in the
lower left inset, six spins on each hexagon form three ferromagnetic pairs.
The spins of each pair are perpendicular to the bond of the pair, and the three
pairs are in all-in or all-out spin structure ($120^\circ$ with each other).
For $K<\frac{3}{2}F$ and $F>-J$ (cyan), the system prefers an in-plane
ferromagnetic state (the lower right inset).

We focus now on the perpendicular ferromagnetic phase.
To obtain the spin wave spectrum, we linearize the LLG equation.
Assume $\mathbf{m}_i=(\delta m_{ix},\delta m_{iy},1)$ being a small
deviation from stable ground state of $\mathbf{m}_0=(0,0,1)$ and substitute
it into the LLG equation \eqref{LLG}. The linearized LLG equation is obtained
by keeping only the linear terms in $\delta m_{ix}$ or $\delta m_{iy}$.
The Bloch theorem guarantees spin wave eigen-solutions of $\delta m_x=X_\beta
e^{i(\mathbf{k}\cdot \mathbf{r}_\beta-\omega t)}$ and $\delta m_y=Y_\beta e^{i(\mathbf{k}
\cdot \mathbf{r}_\beta-\omega t)}$, where $\beta$ denotes sublattices A, B.
The spin wave can be obtained by solving the corresponding
linearized LLG equation \cite{note1}. At K and K$^\prime$ points, the gaps are
$\Delta_g-2\Delta-6\sqrt{3}D$ and $\Delta_g+2\Delta-6\sqrt{3}D$, respectively.
For nonzero DMI $D\neq 0$ and staggered anisotropy $\Delta\neq0$, the system undergoes
topological phase transitions by closing and reopening the gap at one or both valleys.
Fig. \ref{model}(c) shows the topological phase diagram in the $D/J-\Delta/J$ plane
for $K=J$ and $F=0.01J$, which guarantees a perpendicular ferromagnetic ground state
for not too large $\Delta$ and $D$ ($|\Delta|<K$ so that both
$K_\mathrm{A}$ and $K_\mathrm{B}$ are positive, and $|3\sqrt{3}D|<3J+K$).
The Chern number of the conduction
band $\mathcal{C}_c$ is labelled in the figure, and the Chern number of the
valence band $\mathcal{C}_v=-\mathcal{C}_c$ because the sum of the Chern numbers
of all bands must be zero \cite{Niu_book}.  At the phase boundaries one
of the gaps closes, so we obtain the two phase boundaries
$\Delta=\pm(3\sqrt{3}D-\Delta_g/2)$. In the cyan region, the
conduction band has Chern number $-1$, and each valley contributes $-1/2$.
The edge states propagate counterclockwise with respect to $\mathbf{m}$.
If we tune $\Delta$ along $D=0$ ($OO_1$), the gap at K valley closes and reopens by
crossing the phase boundary, and the band Chern number changes from $-1$
to $0$, a transition from topologically nontrivial phase to
trivial phase. If we tune $D$ along $\Delta=0$ ($OO_2$), the gaps at K and
K$^\prime$ close and reopen at the same time, and the band Chern number changes
from $-1$ to $+1$. The system changes from one topologically nontrivial phase
(cyan region where edge states propagate counterclockwise with respect to
$\mathbf{m}$) to another topologically nontrivial phase (pink region where
edge states propagate clockwise with respect to $\mathbf{m}$) \cite{Murakami}. The features
of the phase diagram discussed above do not depend on specific values
of $F$ and $K$ as long as the ground state of the system is the perpendicular
ferromagnetic state.

\begin{figure}[!htb]
\begin{center}
\includegraphics[width=8.5cm]{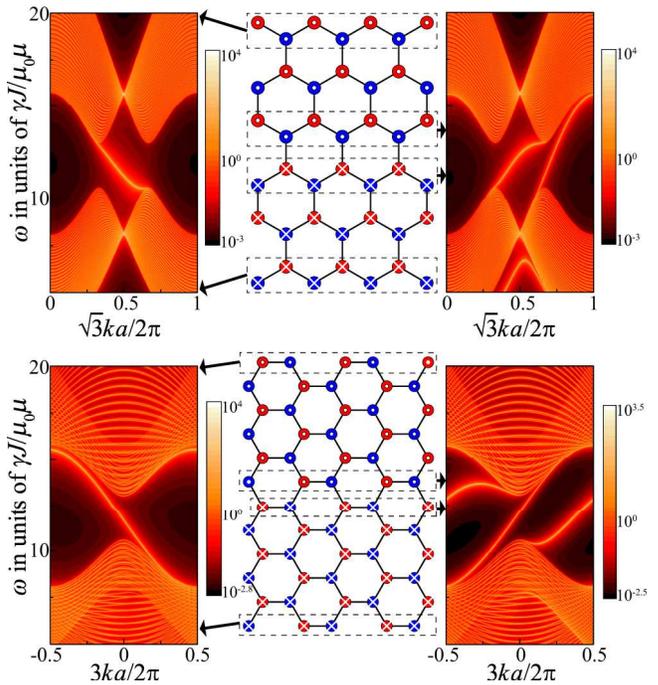}
\end{center}
\caption{The density plots of the spectral function at strip edge or domain
wall (indicated by dashed boxes) for a zigzag strip (upper panel) and an
armchair strip (lower panel). An abrupt domain wall, which separates an
upper domain of $m_z=1$ from a lower domain of $m_z=-1$, sits in the
middle of each strip.}
\label{band}
\end{figure}
\section{CHIRAL EDGE SPIN WAVES}
To reveal the properties of chiral edge spin waves at sample edges and inside a
domain wall, we consider a long strip of zigzag or armchair edges with a domain
wall in the middle, as illustrated in the middle panels of Fig. \ref{band}.
To be specific, we first consider model parameters of $\alpha=10^{-4}$,
$D=\Delta=0$, $K=10J$, and $F=5J$ so that the propagation direction of
topological chiral edge spin waves follows the right-hand rule and domain wall
width $\sqrt{J/K}<1$ is narrow. The spin waves are obtained by solving the eigenvalue
problem $H(k)\Psi =\omega(k) \Psi$, where $H(\mathbf{k})$ is a $4N\times 4N$
matrix with $N=100$ being the number of rows.
The spectral function at $n$th row of the strip is
\begin{equation}
A_n(\omega,k)=-\frac{1}{\pi}\mathrm{Im}\left(\sum_{i=0,1,2,3} G_{(4n-i)(4n-i)}\right),
\end{equation}
where $G_{mn}$ is the matrix element of Green's function $G(\omega,k)=\frac{1}
{\omega-H(k)-i\varepsilon}$ with $\varepsilon$ a small positive number.
The chiral edge spin wave modes in the bulk band gap can be clearly seen from the
density plot of spectral functions as shown in Fig. \ref{band}.  The left panels
are density plots of the spectral functions on the top and the bottom edges
($n=1$ and $n=100$). They perfectly overlap each other, showing
identical dispersion relations of chiral edge spin waves in two domains.
The negative slope of $\omega(k)$ curve says that the spin waves propagate
from the right to the left on both edges at the same speed. The right panels
are the density plots of spectral functions inside the domain wall
($n=50$ and $n=51$). Two edge spin wave modes from the two domains denoted as
$\vert 1\rangle$ and $\vert 2\rangle$ couple with each other inside
the domain wall where they spatially overlap.
The coupling results in two eigenmodes of $(\vert 1\rangle \pm \vert 2
\rangle)/\sqrt{2}$ with different frequencies, one symmetric and the
other antisymmetric with respect to the domain wall central line.
The $\omega(k)$ curves of both modes have positive slopes but with different
values, showing the left-to-right propagation with different velocities.
Importantly, these general features do not depend on sample geometry such as
edge types as shown in Fig. \ref{band}: similar spin wave spectra for
both zigzag strip (upper panel) and armchair strip (lower panel).

\begin{figure*}[!htb]
\begin{center}
\includegraphics[width=17.5cm]{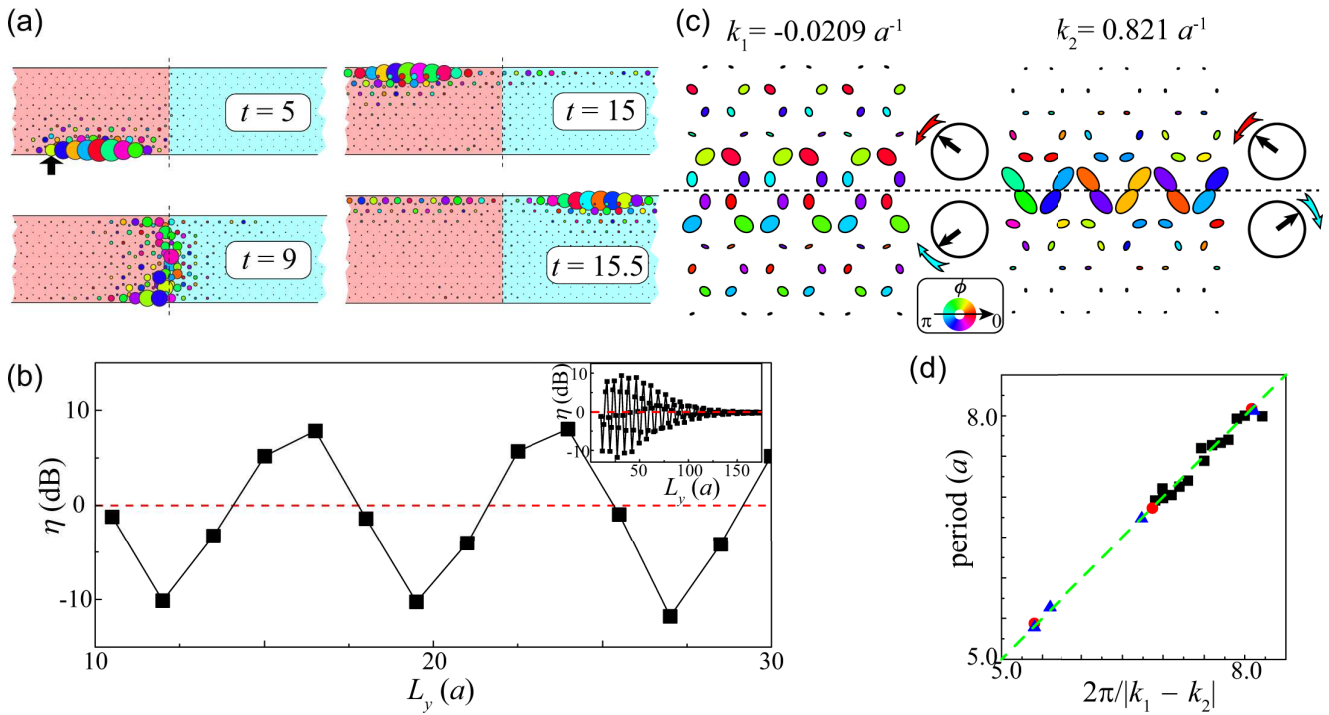}
\end{center}
\caption{
(a) Snapshots of a spin wave before entering ($t=5$), inside
($t=9$), and after leaving ($t=15$) a domain wall of 12 long.
The same spin wave splits differently by a 15 long domain wall ($t=15.5$).
The pink and cyan regions represent respectively $m_z=+1$ and $-1$ domains.
Only portions near the domain wall are shown.
The radius of each circle is proportional to $\sqrt{m_x^2+m_y^2}$,
and the color encodes the azimuthal angles of the spins. (b)
The domain wall length dependence of spin wave power division ratio.
The symbols are simulation results and the red dashed line is $\eta=0$
(for $1:1$ splitting). Inset: $\eta$ approach 0 for large $L_y$.
(c) Spatial distribution of two topologically protected edge spin
waves of $\omega=12$ inside a domain wall parallel to the armchair edges
with wavenumber $k_1=-0.0209$ (left) and $k_2=0.821$ (right).
Insets: spin precession is mirror symmetric (antisymmetric) as
$m_\parallel\rightarrow m_\parallel$ and $m_\perp\rightarrow -m_\perp$
($m_\parallel\rightarrow -m_\parallel$ and $m_\perp\rightarrow m_\perp$
with respect to the domain wall central line for $k_1=-0.0209$ ($k_2=0.821$)
state, here $m_\parallel$ and $m_\perp$ are the magnetization components
parallel and perpendicular to the domain wall.
The symbol shape traces the spin precession trajectories, and the
size of symbols denotes the amplitude of the spin wave at each site.
The azimuthal angles of spins on the lattice at $t = 0$ are
encoded by the symbol colors with the color ring shown in the inset.
(d) The spatial period of the power division ratio from the LLG
simulations (vertical axis) verses the period of beat (horizontal axis).
The green line is $y=x$. The model parameters for squares, circles, and
triangles are respectively $F=5J$ and $K=10J$; $F=6J$
and $K=10J$; and $F=5J$ and $K=9J$. $\alpha$ is set to $10^{-4}$,
and several different frequencies inside the gap are calculated for each
set of parameters.
}
\label{wf}
\end{figure*}
\section{SPIN WAVE BEAMSPLITTER}
After knowing the spectrum of the edge spin waves inside a domain wall,
we investigate numerically how a spin wave beam is split by a domain wall.
We use the same model parameters as those for Fig. \ref{band}, and
consider a strip of $L_x=40\sqrt{3}$ long and $L_y=10.5\sim 180$ wide.
The edges along the $x$ and $y$ directions are respectively
zigzag and armchair types as illustrated in Fig. \ref{wf}(a).
An abrupt domain wall is at $x=L_x/2$. A spin wave beam can be either
injected into one sample edge from outside source or locally generated.
Here a spin wave pulse is locally generated by a microwave field pulse
$\mathbf{h}=0.01[(\cos\omega t)\mathbf{e}_x+(\sin\omega t)\mathbf{e}
_y]$ switched on at $t=0$ for a duration of $\Delta t=5$.
The microwave of $\omega=12$ in the band gap is applied
only at the site marked by the black arrow on the bottom edge.
Typical snapshots of spin wave beam are presented in Fig. \ref{wf}(a).
At $t=5$, the beam is going to enter a 12-long domain wall ($L_y=12$).
At $t=9$, the beam is inside the domain wall. It is clear that spin wave
amplitudes are asymmetric about the domain wall, showing the distortion
of the beam (that is the superposition of two eigenmodes of $\omega=12$)
due to different group velocities of the two spin wave eigenstates.
At $t=15$, the beam leaves the domain wall, and splits into two beams
propagating in opposite directions at the top edge. The intensity of
the two out-going beams are not the same in general. Let us define the
logarithm of the power division ratio as $\eta=\log_{10}\left
(\frac{P_\text{right}}{P_\text{left}}\right)$, where $P_\text{right}$ and
$P_\text{left}$ are the right and left out-going beam powers.
Then $\eta$ depends on the domain wall length, $\eta<0$ for $L_y=12$.
As shown in the fourth panel of Fig. \ref{wf}(a), $\eta$ changes sign
after the same incoming beam passing through a 15 long domain wall.
Fig. \ref{wf}(b) shows that $\eta$ oscillates periodically with $L_y$
for not-too long $L_y$, and approaches to $0$ ($1:1$
splitting) for large $L_y$ as shown in the inset.
Interestingly, the power division ratio does not depend on how far of
the wave source from the domain wall.

To understand this oscillatory behavior, we notice that there are two
topologically protected spin waves for each $\omega$ in the band gap
$(\vert 1\rangle -\vert 2\rangle)/\sqrt{2}$ and $(\vert 1'\rangle +\vert
2'\rangle)/\sqrt{2}$ with different wavenumbers as discussed earlier.
For $\omega=12$, the wavenumbers of the two modes are $k_1=-0.0209$ for
the antisymmetric state (and for $\vert 1\rangle$ \& $\vert 2\rangle$)
and $k_2=0.821$ for the symmetric one (and for $\vert 1'\rangle$ \&
$\vert 2'\rangle$) as indicated by the motion of in-plane components of the
spins on the two sides of the domain wall in the insets of Fig. \ref{wf}(c).
The spatial distributions of the two chiral eigenmodes inside a domain
wall parallel to armchair edges are presented in Fig. \ref{wf}(c).
The mode with $k_2=0.821$ is highly confined around
the domain wall while the mode of $k_1=-0.0209$ is less confined.
The generated spin wave pulse cannot be an eigenmode (an eigenmode must
simultaneously exist in both domains), the spin wave pulse must mainly
be from the two eigenmodes of the same $\omega$ and different $k$.
Since the two modes travel along the same direction at different speed
inside the domain wall as shown earlier, their
superposition generates
a beat pattern with beat wavenumber $k_b=\frac{k_1-k_2}{2}$ when
they are overlapped. Thus, power division ratio should oscillate
with $L_y$ with the period of $\lambda=\left|\frac{2\pi}{2k_b}\right|
=\left|\frac{2\pi}{k_1-k_2}\right|$.  Fig. \ref{wf}(d) shows the period
obtained from LLG simulation (vertical axis) against $\lambda$ from spin
wave spectrum (horizontal axis) for different frequencies and material
parameters. The simulation results coincide with line $y=x$ quite well.
For a given spin wave pulse of time duration $\Delta t$, two eigenmodes
whose speeds inside the domain wall are $v_1$ and $v_2$ will spatially
separate from each other when $L_y$ is longer than $W=\left|\frac{v_1v_2
\Delta t}{v_1-v_2}\right|$. Then spin waves of both eigenmodes leave the
domain wall independently, and their beam power division ratios
should be $1:1$ since they are just single eigenmodes that are the
symmetric or antisymmetric under the permutation of the two edge modes
of the two domains.
This perfectly explains why $\eta$ approach $0$ for large $L_y$.
\begin{figure}[!htb]
\begin{center}
\includegraphics[width=8.5cm]{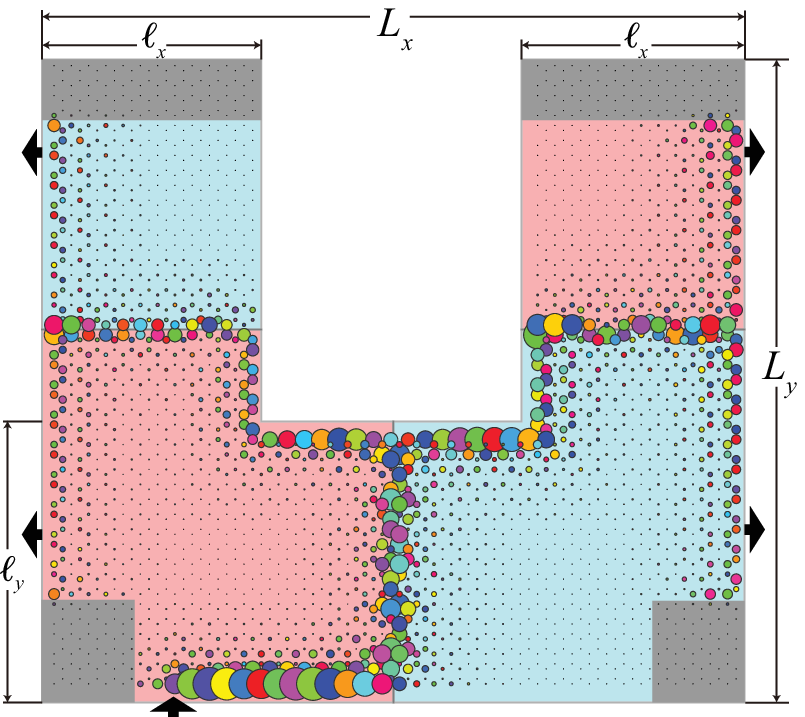}
\end{center}
\caption{
The snapshot of spin wave at $t=65$ under a continuous microwave
excitation of $\omega=12$ at the site marked by the inward arrow
on the lower edge. The outward arrows denote the output signals.
Device geometry, with zigzag edge along the $x$ direction and
armchair edges along the $y$ direction, is $L_x=40\sqrt{3}$,
$L_y=63$, $\ell_y=25.5$, $\ell_x=12\sqrt{3}$. $\alpha=1$ in the
grey areas. }
\label{device}
\end{figure}

The idea of above 1-to-2 SWBS that uses one domain wall to control
topological spin waves can easily be generalized to other 1-to-$n$
SWBSs and spin wave devices by using more domain walls \cite{suppl}.
Fig. \ref{device} illustrates an example of a 1-to-4 SWBS with three
domain walls that separate $m_z=+1$ domains (pink areas) from $m_z=-1$
domains (cyan areas). The grey parts are absorbing areas with a large
damping constant of $\alpha=1$. The figure shows a snapshot of spin
wave pattern at $t=65$ when a microwave field of frequency $\omega=12$
is continuously applied at the site marked by the inward arrow in the
bottom edge. It is clearly shown that a spin wave beam splits into two
beams by the vertically aligned domain wall, and then each of the beam
is further split into two beams by the two horizontally aligned domain
walls in the two arms. The domain wall lengths are designed in such
a way that the spin wave beams are evenly split.
The SWBS can also be used in series to build complicated circuitry.

\section{SPIN WAVE INTERFEROMETER}
\begin{figure*}[!htb]
\begin{center}
\includegraphics[width=17.5cm]{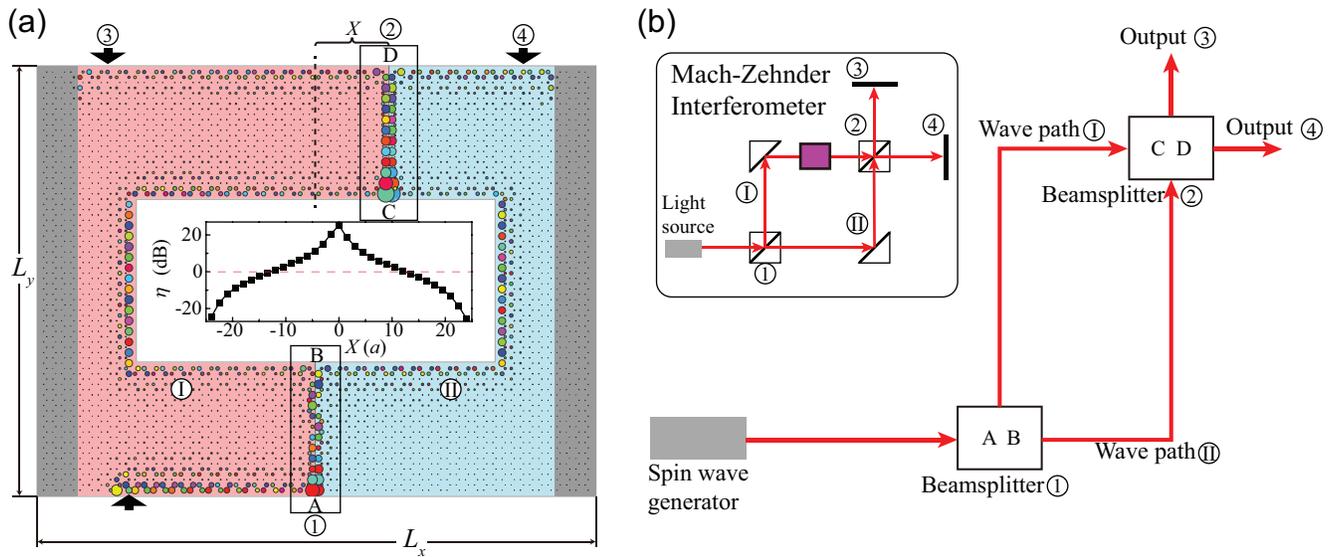}
\end{center}
\caption{
(a) Spin wave beam of $\omega=12$, generated at the site marked
by the inward arrow by a microwave field, splits into beams
\textcircled{\scriptsize{I}} and \textcircled{\scriptsize{II}}
by domain wall AB (the first SWBS).
The two beams recombine in domain wall CD (the second SWBS).
The spin wave pattern at $t=70$ is represented by the size and color
of the symbols that have the same meaning as those in Fig. \ref{wf}.
The device geometry, with armchair edges along the $x$ direction and
zigzag edges along the $y$ direction, is $L_x=90$ and $L_y=40\sqrt{3}$.
The first domain wall of $12\sqrt{3}$ long is placed at $x=L_x/2$, and
splits the in-coming spin wave evenly. An area of $30\times 6\sqrt{3}$
is removed from the center of the device so that the two split spin
wave beams can propagate along the internal boundary. The two beams
recombine at the second domain wall of length $12\sqrt{3}$ at $X$.
The grey parts are absorbing areas with a large damping constant of $\alpha=1$.
Inset: X-dependence of power division ratio $\eta$.
(b) Schematic diagram of the interferometer in (\textbf{a}).
Inset: In the optical Mach-Zehnder interferometer, a light beam
enters optical beamsplitter \textcircled{\scriptsize{1}} and splits into
two beams \textcircled{\scriptsize{I}} and \textcircled{\scriptsize{II}}.
The two beams recombine at the second optical beamsplitter
\textcircled{\scriptsize{2}}. The outputs \textcircled{\scriptsize{3}}
and \textcircled{\scriptsize{4}} depend on the interference of
two beams at beamsplitter \textcircled{\scriptsize{2}}.
}
\label{device2}
\end{figure*}

Fig. \ref{device2}(a) is a proposal of a Mach-Zehnder-type spin wave
interferometer with two domain walls separating a left $m_z=+1$ domain
(pink area) from a $m_z=-1$ domain (cyan area). A spin wave beam of
$\omega=12$ generated at the site marked by the inward arrow enters
the first domain wall of length AB. The beam splits evenly to beams
\textcircled{\footnotesize{I}} and \textcircled{\footnotesize{II}}
by the SWBS as explained earlier. After travel certain distances, the
two beams recombine in the second SWBS (domain wall of length CD).
Spin waves can go to either \textcircled{\footnotesize{3}} or
\textcircled{\footnotesize{4}} \cite{suppl}. Their intensities should
depend on the interference of two beams inside the second domain wall.
Fig. \ref{device2}(a) shows the snapshot of spin wave pattern at
$t=70$ when a spin wave beam is emitted into the device at $t=0$.
The above precess can be schematically represented by the diagram in Fig.
\ref{device2}(b) that is exactly the same as the diagram for the optical
Mach-Zehnder interferometer shown in the inset of Fig. \ref{device2}(b).
Of course, instead of light and two optical beamsplitters in an optical
Mach-Zehnder interferometer, we have spin wave and two SWBSs here.
In our interferometer, the relative phase of the two interfered spin
waves can be tuned, for example, by placing the second domain wall at
different positions or by changing the length of the second domain wall.
The interference pattern is reflected by the power division ratio after
the spin wave beam comes out of the second domain wall
(beamsplitter \textcircled{\footnotesize{2}}).
The inset of Fig. \ref{device2}(a) is the position ($X=0$ when two
domain walls align along the same vertical line) dependence of power
division ratio $\eta$ of the second domain wall of $12\sqrt{3}$ long.
This is in contrast to the simple SWBS discussed earlier whose power
division ratio does not depend on the location of the domain wall.
Since the position of the second domain wall can be
controlled by magnetic fields or electric current/fields, this
device can also be used as a tunable SWBS or a spin wave multiplexer.

\section{DISCUSSION AND CONCLUSION}
The results reported here do not depend on the details of the model
as long as the system supports topologically protected unidirectional
spin waves \cite{TM1,TM2,TM3}. In practical applications,
one would like to use magnetic materials with low damping such as YIG
whose damping can be as low as $\alpha<10^{-5}$ so that spin decay
length is about a millions of wavelength \cite{magnonics2,Kajiwara}.
We consider abrupt domain walls with strong anisotropy here, and it
should be interesting to also consider the case with wide domain walls.
The spin wave interferometer shows a lot of similarities to the
optical Mach-Zehnder interferometer. Although we study spin waves at
classical level in our model, it is also possible to repeat the
study at quantum level so that one can investigate interesting
quantum phenomena such as magnonic Hong-Ou-Mandel effect \cite{HOM}.

Due to the unidirectional property of the edge spin waves, basic
magnonic components such as spin wave diodes, circulators, and gyrators
can also be designed utilizing topological magnetic materials.
Thus, our proposal is possible to realize programmable on-chip
integrated circuits, a magnonic analogy of Silicon-based electric
integrated circuits with the advantage of reconfigurablity.
It allows one to draw, erase, and redraw a complicated spin wave circuit
on a magnetic plate as one wishes since the domain configuration
can be manipulated by magnetic field and/or electric current/field.
A domain configuration can be fixed by an antiferromagnetic
layer through exchange bias effect \cite{ex_bias} if it is needed.
Furthermore, the performance of the devices and circuitry can be
effectively controlled and tuned by magnetic fields and electric fields
through the control of material properties and domain wall properties.
There are different ways to experimentally realize the devices.
Any system that supports topologically protected unidirectional
spin waves can be used to build the proposed circuity.
For example, in $\text{Lu}_2\text{V}_2\text{O}_7$ with a pyrochlore structure
\cite{Nagaosa} or Cu[1,3-bdc] with a Kagome structure \cite{Kagome}, there
are already experimental evidences for the existence of topological spin
wave edge states. To realize the model studied above, materials with strong
pseudodipolar exchange interaction and/or DMI are needed, and heavy metal
compounds with strong spin-orbit coupling could be a direction to look for.

In conclusion, we demonstrated the controlled spin wave propagation
using topologically protected edge states.
Reconfigurable spin wave beamsplitters and
spin wave interferometers were designed and studied. The power division
ratio of the spin wave beamsplitter oscillates with the domain wall
length due to the interference of two
spin waves in two eigenmodes of the same frequency and different wavenumbers.

\section{ACKNOWLEDGMENTS}
This work was supported by National Natural Science
Foundation of China (Grant No. 11374249) and Hong Kong
RGC (Grant No. 16301115 and 16301816). X.S.W acknowledge
support from UESTC and China Postdoctoral Science Foundation
(Grant No. 2017M612932).


\begin{thebibliography}{99}
\bibitem{note1}By defining $\psi^\pm_\beta=(X_\beta\pm Y_\beta)/\sqrt{2}$ ($\beta=$ A, B),
the column vector $\Psi=
(\psi^{+}_\mathrm{A},\psi^{-}_\mathrm{A},\psi^{+}_\mathrm{B},
\psi^{-}_\mathrm{B})^\mathrm{T}$ satisfies the eigenvalue problem
$gH(\mathbf{k})\Psi=\omega(\mathbf{k})\Psi$, where $H(\mathbf{k})$ is a $4\times 4$ Hermitian
matrix depending on the model parameters $K/J$, $F/J$, $\Delta/J$, $D/J$, and
$g=\sigma_0\bigotimes\sigma_3$ (with $\sigma_0$ being the $2\times2$
identity matrix and $\sigma_3$ the Pauli matrix).
\end{thebibliography}
\end{document}